\documentstyle[aps,epsf,floats]{revtex}  

\begin{document}        

\baselineskip 14pt
\title{The Status of MSSM Higgs Boson Searches at LEP}
\author{Andy Hocker}
\address{University of Chicago}
%
\maketitle              

\begin{abstract}        
The most recently available results from searches conducted by the
four LEP experiments at 189 GeV center-of-mass energy for Higgs bosons of
the Minimal Supersymmetric
Standard Model (MSSM) are presented.  No evidence for a signal has been
observed, and the null result is used by the experiments, both individually
and collectively, to exclude regions of the MSSM parameter space and to
set lower limits on Higgs boson masses at 95\% confidence level in
constrained MSSM scenarios.
\
\end{abstract}   	

\section{Introduction}               

The problem of the origin of electroweak symmetry breaking in
fundamental theories of particle physics is often solved by
invoking a Higgs mechanism, where the symmetry is broken by
the introduction of one or more scalar field doublets, which in
turn give rise to the existence of physical neutral scalar particles
called Higgs bosons.  Unfortunately the Higgs masses are left
as free parameters of the theory.  Current fits to precision
electroweak data, for example, can allow a Standard
Model Higgs boson ($\mathrm{H}^0_{\mathrm{SM}}$) mass of up to 262
GeV/$c^2$~\cite{RefEWfit}.

A common feature of supersymmetric extensions to the SM,
however, is the prediction of the existence of a relatively
light Higgs boson.  In particular, the Minimal Supersymmetric
Standard Model (MSSM) predicts the existence of five Higgs bosons:
two neutral and CP-even ($\mathrm{h}^0$ and $\mathrm{H}^0$, with
$m_{\mathrm{h}^0} < m_{\mathrm{H}^0}$ by definition), one
neutral and CP-odd ($\mathrm{A}^0$), and two charged ($\mathrm{H}^{\pm}$).
At tree level, $m_{\mathrm{h}^0}$ is predicted to be less than
$m_{\mathrm{Z}^0}$; however, radiative corrections depending strongly
on the top quark mass and mixing in the MSSM's stop sector significantly
alter this relation.  The stop-mixing terms in turn depend on a number
of unknown MSSM parameters, but an upper bound on $m_{\mathrm{h}^0}$
can still be set at around 130 GeV/$c^2$~\cite{Refhub} independent of
the choice of these parameters.  A substantial fraction of this
mass range can be explored at LEP2.

In 1998, the four LEP experiments (ALEPH, DELPHI, L3, and OPAL) each collected
over 150 $\mathrm{pb}^{-1}$ of $\mathrm{e}^+\mathrm{e}^-$ collision
data at $\sqrt{s} \approx 189$ GeV.  The individual
experimental results presented
here are those most recently available, and are in most cases based on
only a partial sample of the 1998 data.  Therefore they should be
regarded as very preliminary.

\section{Production and Decay of Higgs Bosons at LEP2}

\begin{figure}[ht]	
\centerline{\epsfxsize 5.0 truein \epsfbox{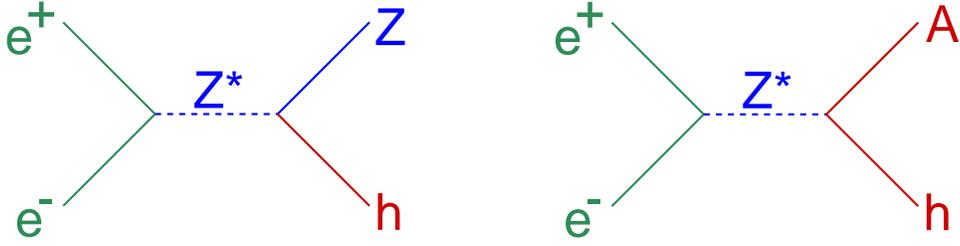}}   
\vskip -.2 cm
\caption[]{
\label{FigFeyn}
\small Feynman diagrams for the $\mathrm{h}^{0}\mathrm{Z}^{0}$
Higgsstrahlung and $\mathrm{h}^{0}\mathrm{A}^{0}$ pair-production
processes.}
\end{figure}

Figure~\ref{FigFeyn} shows the two dominant MSSM Higgs production
mechanisms at LEP2 (``Higgsstrahlung'' and ``pair-production'').
Their cross-sections are given by
\begin{eqnarray}
& &
\sigma_{\mathrm{hZ}} = \sin^2(\beta-\alpha)
\sigma ^{\mathrm{HZ}}_{\mathrm{SM}} \\
& &
\sigma_{\mathrm{hA}} = \cos^2(\beta-\alpha)\bar{\lambda}
\sigma ^{\nu\bar{\nu}}_{\mathrm{SM}}
\end{eqnarray}
where $\tan\beta$ is the ratio of the VEV's of the two neutral CP-even
Higgs fields, $\alpha$ is the mixing angle between them, and $\bar{\lambda}$
is a kinematic factor.  A complementarity between the two processes can
be seen in the appearance of the $\cos^2$ and $\sin^2$ terms.

The Higgsstrahlung process is the radiation of a Higgs by a virtual
$\mathrm{Z}^0$ that subsequently goes on-shell.  This process is
favored in the low $\tan\beta\ (\approx 1-2)$ regime, where it is
almost indistinguishable from SM Higgs production and decay.  Hence,
the SM Higgs searches are ``recycled;'' we look for a pair of b-jets
or $\tau$ leptons recoiling from a fermion-antifermion pair with the
mass of the $\mathrm{Z}^0$ (the Higgs couples preferentially to
high-mass particles; at LEP2 the b and $\tau$ are the most massive
kinematically-allowed decay products).

The second process is the pair-production of the $\mathrm{h}^0$
and $\mathrm{A}^0$ from the decay of a virtual $\mathrm{Z}^0$.
This is the favored process for large $\tan\beta\ (>15)$ and gives
rise to final states with four heavy fermions.  Therefore dedicated
searches for the $\mathrm{b\bar{b}b\bar{b}}$ and
$\mathrm{b\bar{b}}\tau^{+}\tau^{-}$ final states have been developed.

A third set of topologies can exist in a small subset of the MSSM parameter
space where $m_{\mathrm{h}^0} > 2m_{\mathrm{A}^0}$.  In this case
$\mathrm{h}^0 \rightarrow \mathrm{A}^0\mathrm{A}^0$ may be the
dominant decay.  The SM Higgs searches still retain a respectable
efficiency for this decay when it occurs within the Higgsstrahlung
process; in the pair-production process, OPAL performs a dedicated
search for the $\mathrm{b\bar{b}b\bar{b}b\bar{b}}$ final state.

\section{Experimental Approach}

\begin{figure}[ht]	
\centerline{\epsfxsize 5.0 truein \epsfbox{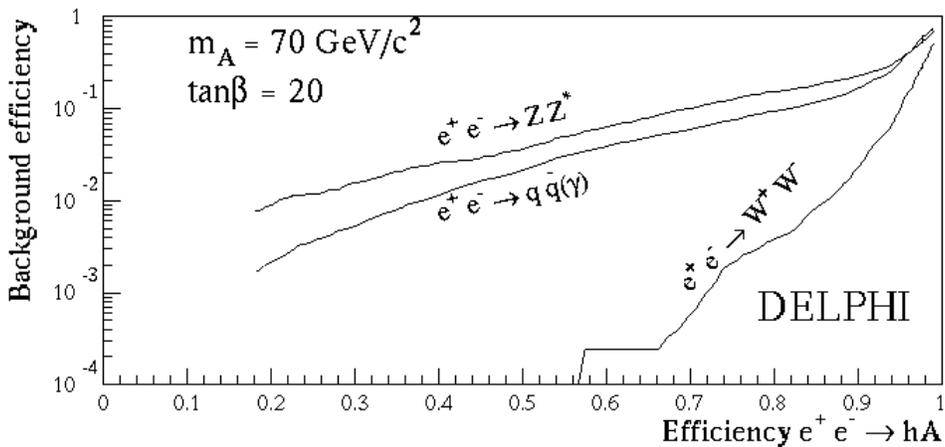}}   
\vskip -.2 cm
\caption[]{
\label{Figbtag}
\small DELPHI b-tagging efficiency for Higgs background processes versus
signal efficiency.}
\end{figure}

The predominance of b quarks in the Higgs final states makes b-tagging
one of the most important tools in Higgs searching.  To that end, all
the LEP detectors were equipped with silicon microvertex detectors for
precision secondary vertexing of the charged tracks produced in long-lived
B hadron decays.  The lifetime information is combined with other
discriminating variables to yield
high-efficiency b identification.  As an example of the performance of
these algorithms, Figure~\ref{Figbtag} shows the DELPHI b-tagging efficiency
with respect to the dominant Higgs backgrounds as a function of the
efficiency for $\mathrm{e}^+\mathrm{e}^-\rightarrow\mathrm{h}^0\mathrm{A}^0$.
For a signal efficiency of 70\%, the background from $\mathrm{Z}^0$-pairs
and QCD is reduced by an order of magnitude, and in the case of the b-less
W-pair decays, several orders of magnitude.

Since the b-tagging is so crucial to the analyses, it is important that
its performance is well-understood.  For example, OPAL cross-checks Monte
Carlo efficiency/fake-rate predictions with high-energy data samples such
as radiative returns to the $\mathrm{Z}^0$ pole and semileptonic W-pair decays.
Checks like these ensure the robustness of the analyses performed in the
channels described below.

\subsection{The 4b Channel}

B-tagging information and kinematic quantities are combined to discriminate
signal-like $\mathrm{b\bar{b}b\bar{b}}$ events from background, usually
via sophisticated algorithms such as multivariate relative likelihoods
or artificial neural networks.
A summary of the individual experiments' performance in this channel
is given in Table~\ref{Tab4b}.
The signal efficiency typically ranges from 50-65\% (the exact number
depends on what Higgs masses are under consideration), while the accepted
background cross-section is reduced to tens of femtobarns.  No significant
excess is seen in the data.

 \begin{table}
 \caption{\label{Tab4b} Performance summary of the LEP 4b analyses.}
 \begin{tabular}{cdddc} 
     &    analyzed ${\mathcal L}(\mathrm{pb}^{-1})$ &
                 number of events & number of events &   typical  \\
     & at 189 GeV  & expected & observed & efficiency \\
\tableline
ALEPH  & 35.7 & 1.9 & 3 & 60\% \\
DELPHI & 158.0 & 11.1 & 11 & 65\% \\
L3\tablenote{Specific numbers from L3 were not available at conference time.
This is also the case in Table~\ref{Tabtau}.} & 32.9 \\
OPAL   & 151.0 & 7.0 & 12 & 50\% \\
 \end{tabular}
 \end{table}

\subsection{The Tau Channel}

To search for $\mathrm{b\bar{b}}\tau^+\tau^-$ events, b-tagging and
kinematic quantities are again combined, this time in addition to
tau-tagging schemes of varying complexities.  Table~\ref{Tabtau}
summarizes the performance of the experiments' analyses in this channel.
Efficiencies are somewhat lower than in the 4b channel, yet the accepted
background cross-section can be reduced to a few femtobarns.  Again, no
excess is observed in the data.

 \begin{table}
 \caption{\label{Tabtau} Performance summary of the LEP tau analyses.}
 \begin{tabular}{cdddc} 
     &    analyzed ${\mathcal L}(\mathrm{pb}^{-1})$ &
                 number of events & number of events &   typical  \\
     & at 189 GeV  & expected & observed & efficiency \\
\tableline
ALEPH  & 35.7 & 0.1 & 0 & 30\% \\
DELPHI & 158.0 & 0.6 & 0 & 20\% \\
L3     & 32.9 \\
OPAL   & 149.4 & 4.8 & 5 & 40\% \\
 \end{tabular}
 \end{table}

\subsection{The 6b channel}

The presence of 6 b-flavored jets in the $\mathrm{b\bar{b}b\bar{b}b\bar{b}}$
final state is a striking enough topology that a simple analysis can
be afforded.  Cuts on the charged multiplicity of the event, jet-finding
resolution parameters, and b-tags are sufficient to achieve good efficiency
with a manageable background.  OPAL expects 7.3 background events and observes
eight in 151 $\mathrm{pb}^{-1}$ of analyzed 189 GeV data while retaining signal
efficiencies around 60\%.  It is worth noting that no explicit mass
reconstruction is done in this
channel due to the combinatorics involved in reconstructing the six jets to
two bosons.

\section{Interpretation and Results}

The non-observation of Higgs production allows us to rule out\footnote{All
exclusions and mass limits presented here are at 95\% confidence level.} MSSM
scenarios and Higgs masses yielding observable cross-sections.  In its
most general form the MSSM has more than one hundred free parameters,
but many of them have no impact on Higgs phenomenology.  Bearing that
in mind, results are interpreted within a constrained MSSM
where unification of the sfermion
masses at the GUT scale is assumed, as well as unification of the sfermion
tri-linear couplings and gaugino masses at the electroweak scale.

\subsection{Benchmark Scan}

\begin{figure}[ht]	
\centerline{\epsfxsize 3.0 truein \epsfbox{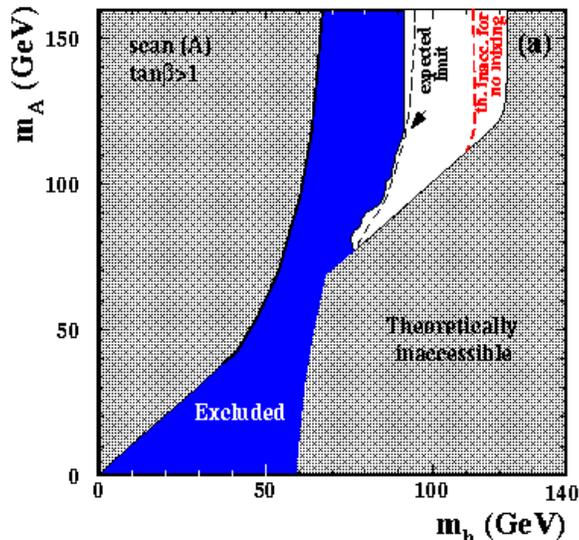}}   
\vskip -.2 cm
\caption[]{
\label{FigBmhma}
\small Regions of the $m_{\mathrm{h}^0}-m_{\mathrm{A}^0}$ plane excluded
by experiment (dark shading) and theory (light
shading) for $\tan\beta > 1$ (preliminary OPAL 189 GeV benchmark scan).}
\end{figure}

\begin{figure}[ht]	
\centerline{\epsfxsize 3.0 truein \epsfbox{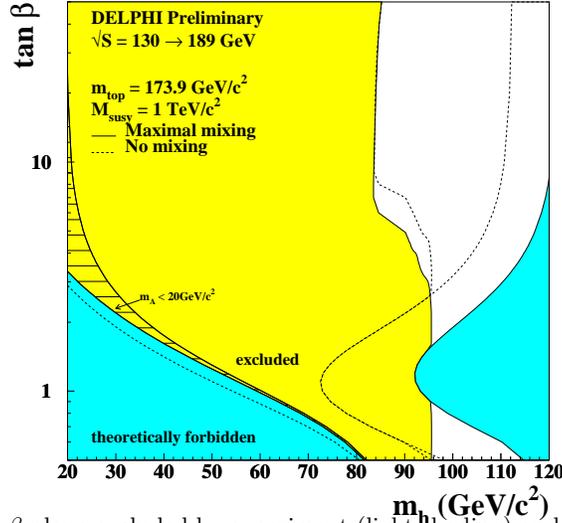}}   
\vskip -.2 cm
\caption[]{
\label{FigBmhtb}
\small Regions of the $m_{\mathrm{h}^0}-\tan\beta$ plane excluded by
experiment (light shading) and ``maximal stop-mixing'' theory (dark
shading) in the benchmark scan.  The long dotted line shows the larger
excluded area if only minimal stop-mixing is considered.}
\end{figure}

Following a prescription for a benchmark set of MSSM parameters set forth
in~\cite{RefYbook},
a large number of possible
values of $\tan\beta$ and the running $\mathrm{A}^0$ mass are scanned
while keeping the soft SUSY-breaking masses, the top quark mass\footnote{The
experimental uncertainty on $m_t$ combined with the Higgs sector's strong
sensitivity to this parameter makes $m_t$ ``quasi-free.''}, and the
supersymmetric Higgs mass parameter $\mu$ fixed.  In addition two possible
mixings in the stop sector are considered (minimal and maximal).
Figure~\ref{FigBmhma} shows an example of the results of this scan in the
$m_{\mathrm{h}^0}-m_{\mathrm{A}^0}$ plane for $\tan\beta > 1$.  The white
area shows the only region that is not excluded by either theory or
experiment.  Lower limits on $m_{\mathrm{h}^0}$ and
$m_{\mathrm{A}^0}$ can then be read off from the lower left corner of this
region; these limits are summarized in Table~\ref{TabBlim}.  Another
projection of this scan, this time in the $m_{\mathrm{h}^0}-\tan\beta$
plane~(Figure~\ref{FigBmhtb}), shows that a range of low $\tan\beta$ can be
excluded {\emph even with maximal stop-mixing.}  This exclusion has only
become available with the 189 GeV data.  It should be noted, however,
that this exclusion vanishes using a top mass $2\sigma$ larger than its
measured central value. 
 
 \begin{table}
 \caption{\label{TabBlim} Lower mass limits for $\tan\beta > 1$,
based on varying amounts of analyzed data at 189 GeV.}
 \begin{tabular}{cdddd} 
  & ALEPH\tablenote{Results based on hA search only}  & DELPHI & L3 & OPAL \\
\tableline
$m_{\mathrm{h}^0}$ (GeV/$c^2$) & 79.7 & 83.5 & 74 & 76 \\
$m_{\mathrm{A}^0}$ (GeV/$c^2$) & 79.7 & 84.5 & 74 & 77 \\
 \end{tabular}
 \end{table}

\subsection{General Scan}

\begin{figure}[ht]	
\centerline{\epsfxsize 3.0 truein \epsfbox{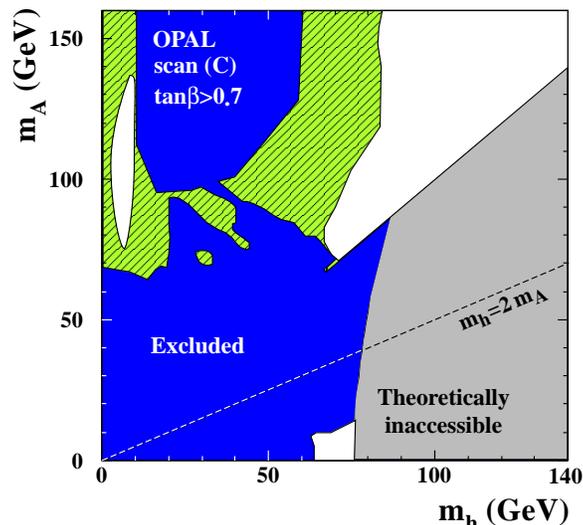}}   
\vskip -.2 cm
\caption[]{
\label{FigGmhma}
\small Results of the OPAL 183 GeV general scan in the
$m_{\mathrm{h}^0}-m_{\mathrm{A}^0}$ plane.  Unexcluded regions are
shown in white.}
\end{figure}

A more general interpretation is obtained by ALEPH, DELPHI, and OPAL
by releasing all the parameters fixed in the benchmark scan.  However,
large areas of this enormous parameter space can be excluded on physical
grounds, such as requiring the absence of charge and color-breaking
minima in the MSSM Lagrangian and requiring neutralino and stop masses
that are unexcluded by direct searches.  An example of this scan in the
$m_{\mathrm{h}^0}-m_{\mathrm{A}^0}$ plane is shown in Figure~\ref{FigGmhma}
(this scan only uses data taken up to and including 1997's 183 GeV run).
Some conclusions drawn from these scans are that absolute mass limits
are generally 5-10 GeV/$c^2$ worse than those derived from the benchmark scan,
and that the limit-weakening parameter sets constitute 0.01-0.1\% of those
scanned.  These sets are usually characterized by a small
$\sin^2(\beta-\alpha)$ (such that $\mathrm{h}^0\mathrm{Z}^0$ production
is heavily suppressed) and an $\mathrm{A}^0$ out of LEP2's kinematic reach.

\subsection{LEP-wide Combinations}

It can be seen that the individual experiments' mass limits are well
below the kinematic limit; therefore substantial gains can be made by
pooling the luminosities of the four experiments.  This is not a
straightforward procedure; the LEP Higgs Working Group has investigated
four statistical procedures for combining the individual results, described
in~\cite{RefLEP}.  This combination has been done within the context
of the benchmark scan for all data up to and including 1997's 183 GeV run.
Quantitative results from this combination include an exclusion
of $\tan\beta$ in the range 0.8-2.1 for minimal stop-mixing and
$m_t=175$ GeV/$c^2$, and lower mass limits on $m_{\mathrm{h}^0}$ and
$m_{\mathrm{A}^0}$ of 78.8 and 79.1 GeV/$c^2$, respectively.  These mass
limits represent a gain of about 10 GeV/$c^2$ with respect to the experiments'
individual 183 GeV results.

\section{Charged Higgs Bosons}

\begin{figure}[ht]	
\centerline{\epsfxsize 3.0 truein \epsfbox{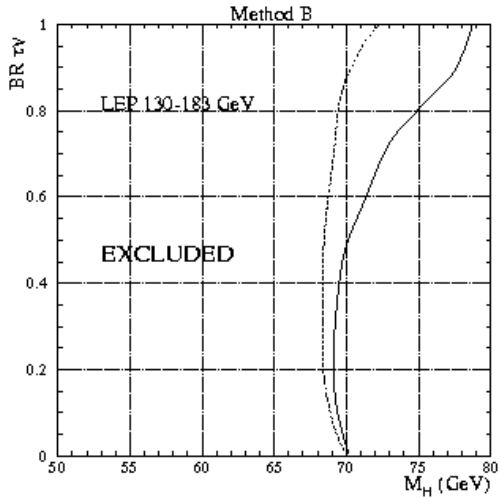}}   
\vskip -.2 cm
\caption[]{
\label{FigCh}
\small LEP-wide charged Higgs lower mass limit as a function of
$\mathrm{BR}(\mathrm{H}^\pm \rightarrow \tau\nu_\tau)$.  The dotted line
represents the observed limit; the solid line is the average expected limit
obtained from a large number of background-only experiments.}
\end{figure}

Charged Higgs bosons could be produced at LEP2 from the decay of a virtual
$\mathrm{Z}^0$ into a $\mathrm{H}^+\mathrm{H}^-$ pair.  Most MSSM scenarios
predict a charged Higgs mass that puts it out of reach of LEP2; however,
light charged Higgses can exist in some more general two-Higgs-doublet models.
Searches for $\mathrm{H}^+\mathrm{H}^-$ have been performed at LEP2 in the
hadronic ($\mathrm{c\bar{s}\bar{c}s}$), semileptonic
($\mathrm{cs}\tau\nu_\tau$) and leptonic ($\tau^+\nu_\tau\tau^-\bar{\nu_\tau}$)
channels.  This search is experimentally quite challenging due to the large
background from W-pair decays with topologies nearly identical to the signal.
No evidence for a signal has been observed in the data, and lower limits are
placed on $m_{\mathrm{H}^\pm}$ as a function of the
branching ratio to $\tau\nu_\tau$ and assuming
$\mathrm{BR}(\tau\nu_\tau)+\mathrm{BR}(\mathrm{q\bar{q}'})=1$.
The charged Higgs LEP-wide combination is a project that is still in its
infancy, but the result of one preliminary combination (again only up to
183 GeV) is shown in Figure~\ref{FigCh}, which gives a lower limit of
68 GeV/$c^2$, representing a gain of about 10 GeV/$c^2$ with respect to the
experiments' individual 183 GeV limits.

\section{Conclusion}

Despite active searching, the LEP experiments have yet to find any
evidence for MSSM Higgs boson production.  They have combined their
results from data taken at center-of-mass energies from 91 to 183 GeV to
place lower benchmark limits on $m_{\mathrm{h}^0}$ and
$m_{\mathrm{A}^0}$ of 78.8 and 79.1 GeV/$c^2$, respectively.  In addition,
they exclude the range $0.8 < \tan\beta < 2.1$ for minimal stop-mixing
and $m_t = 175$ GeV/$c^2$.  However, the experiments' individual results
from 189 GeV have already begun to supersede the combined results; for
example, DELPHI's lower limits on $m_{\mathrm{h}^0}$ and $m_{\mathrm{A}^0}$ are
83.5 and 84.5 GeV/$c^2$, respectively.  In addition, exclusions of $\tan\beta$
with $m_t=175$ GeV/$c^2$ are becoming available for \emph{any} stop-mixing
scenario.

The final years of LEP2 running promise to be exciting ones as the
center-of-mass energy is pushed up to 200 GeV and possibly beyond.  Either
the Higgs boson will be discovered, or we will be able to severely constrain
the numerous possible manifestations of the MSSM.


\end{document}